\documentclass[prb,aps, twocolumn,floatfix,showpacs,showkeys]{revtex4-1}

\usepackage{graphicx}
\usepackage{epstopdf}

\usepackage{color}

\usepackage[utf8]{inputenc}
\usepackage{xcolor}

\begin{document}

\title{Correlation between charge density waves and antiferromagnetism \\in Nd$_{1-x}$Gd$_x$NiC$_2$  solid solution}

\author{Marta Roman, Tomasz Klimczuk, Kamil K. Kolincio} 

\affiliation{Faculty of Applied Physics and Mathematics, Gdansk University of Technology,
Narutowicza 11/12, 80-233 Gdansk, Poland}

\begin{abstract}
We report a study on the evolution of a charge density wave and antiferromagnetism in the series of the polycrystalline solid solution Nd$_{1-x}$Gd$_x$NiC$_2$ (0 $\leq$ x $\leq$ 1) by means of magnetic and transport properties measurements. The experimental results reveal the violation of the de Gennes law and a strong correlation between the Peierls, Néel and Curie-Weiss temperatures, which strongly suggests a cooperative interaction between the charge density wave state and antiferromagnetism due to Fermi surface nesting enhancement of the RKKY interaction. We also find that, the obtained results for the Nd$_{1-x}$Gd$_x$NiC$_2$ (0 $\leq$ x $\leq$ 1) series overlap with the $T_{CDW}$ trend line in the phase diagram for RNiC$_2$ family.
\end{abstract}

\maketitle

\section{Introduction}
Quasi low-dimensional systems offer a large variety of unique physical properties such as charge density wave (CDW) or spin density wave (SDW) instabilities \cite{Monceau2012, Gruner2000, Gruner1988}. The low dimensionality of the electronic structure is also seen as an important ingredient of high temperature superconductivity (SC) and the charge density wave state has been found to be universal feature in the phase diagrams of the cuprate superconductors family\cite{Leyraud2013, Caprara2017, Thampy2017, Zhou2017, Caplan2017, Kawasaki2017, Laliberte2018, Cyr2018}.
For this reason, the interplay between various types of ordering such as CDW, SC and magnetism is a central issue in solid state physics \cite{Fawcett1988, Jacques2014, Xu2009, Chang2012, Chang2016, Graf2004, Winter2013, Murata2015, Gruner2017}. The rich phase diagram of the low-dimensional rare earth nickel dicarbides RNiC$_2$ in which various ground states such as ferromagnetic (FM), antiferromagnetic (AFM), superconducting and charge density wave states have been reported so far, makes the members of this family appropriate candidates for the investigation of the relations between numerous types of ordering. The ground state of the members of this family depends on the rare-earth metal component denoted by R.
 LaNiC$_2$ is a noncentrosymmetric superconductor below T$_{sc}$=2.7 K \cite{Wiendlocha2016, lee_superconductivity_1996, Pecharsky1998, Quintanilla2010, Landaeta2017}, SmNiC$_2$ undergoes a ferromagnetic transition at $T_C$=17.5 K \cite{Onodera1998} and the rest of the compounds (apart from Pr where a weak magnetic anomaly is observed \cite{Onodera1998, Kolincio2017}) order antiferromagnetically\cite{Onodera1995, Onodera1998, Kotsanidis1989}. In this system, the magnetic order originates entirely from the 4$f$ electrons of the rare earth ions R$^{3+}$ acting as local magnetic moments interacting through the Ruderman-Kittel-Kasuya-Yosida (RKKY) interaction. 
The CDW state has been found for most of the members of the RNiC$_2$ family (R = Pr - Lu) with the temperature ranging from 89 K for PrNiC$_2$ to 463 K for LuNiC$_2$ \cite{Murase2004, Laverock2009, Wolfel2010, Sato2010, Ahmad2015, Michor2017, Roman2018}.
Remarkably the Peierls temperature $T_{CDW}$ and the lock-in transition temperature $T_1$ have been found to scale linearly with the unit-cell volume for R ranging from Sm to Lu \cite{Roman2018}. This effect has been tentatively attributed to the evolution of the Fermi surface (FS) topology resulting in the modification of the nesting conditions. Interestingly, the Peierls temperature for Nd and Pr bearing compounds deviates from the linear trend observed for the rest of the family.

The CDW in RNiC$_2$ has been found to interact with the magnetic state. For SmNiC$_2$, the Peierls instability is completely suppressed below the Curie temperature \cite{Shimomura2009, Hanasaki2012, Lei2017, Kim2012}, in contrast with PrNiC$_2$, where the magnetic anomaly has been found to have a constructive impact on the nesting properties\cite{Yamamoto2013, Kolincio2017}. In the compounds showing antiferromagnetic ground state, a CDW partially survives below the Néel temperature\cite{Yamamoto2013, Kolincio20161, Kolincio2017, Hanasaki2017}. Recently Hanasaki et al. \cite{Hanasaki2017} suggested that the AFM order originates from the cooperative effect involving a CDW and spin oscillations. These reports inspired us to explore the evolution of a CDW instability and magnetism on the path between NdNiC$_2$ and GdNiC$_2$, both exhibiting an antiferromagnetic ground state and  standing on opposite sides of the deviation from the linearity on the RNiC$_2$ phase diagram.
\\
\indent
In this paper we report a detailed investigation on the solid solution Nd$_{1-x}$Gd$_x$NiC$_2$ (0 \(\le\) x \(\le\) 1) by means of powder X-ray diffraction, AC and DC magnetic susceptibility and electrical resistivity. The results were discussed with a particular emphasis on the interrelationship between a CDW state and antiferromagnetic ordering.

\section{Experimental}
The series of the polycrystalline Nd$_{1-x}$Gd$_x$NiC$_2$ solid solutions for Gd concentration 0 \(\le\) x \(\le\) 1  were prepared by arc-melting of the proper amounts of pure elements: Ni (3N), C (5N), and Nd (3N) and Gd (3N) in a high purity argon atmosphere with a zirconium button used as an oxygen getter. To compensate for the loss during the arc-melting process additional amounts of rare earth metals (\(\approx\)2\%) and carbon (\(\approx\)3\%) were used. All samples were turned over and remelted four times on water-cooled copper hearth in order to obtain good homogeneity. All the samples obtained from arc-melting were wrapped in tantalum foil, placed in an evacuated quartz tube, annealed at 900$^o$C for 12 days and cooled down to room temperature by quenching in cold water. 
  \begin{figure*}[ht]
\includegraphics[angle=0,width=2.1\columnwidth]{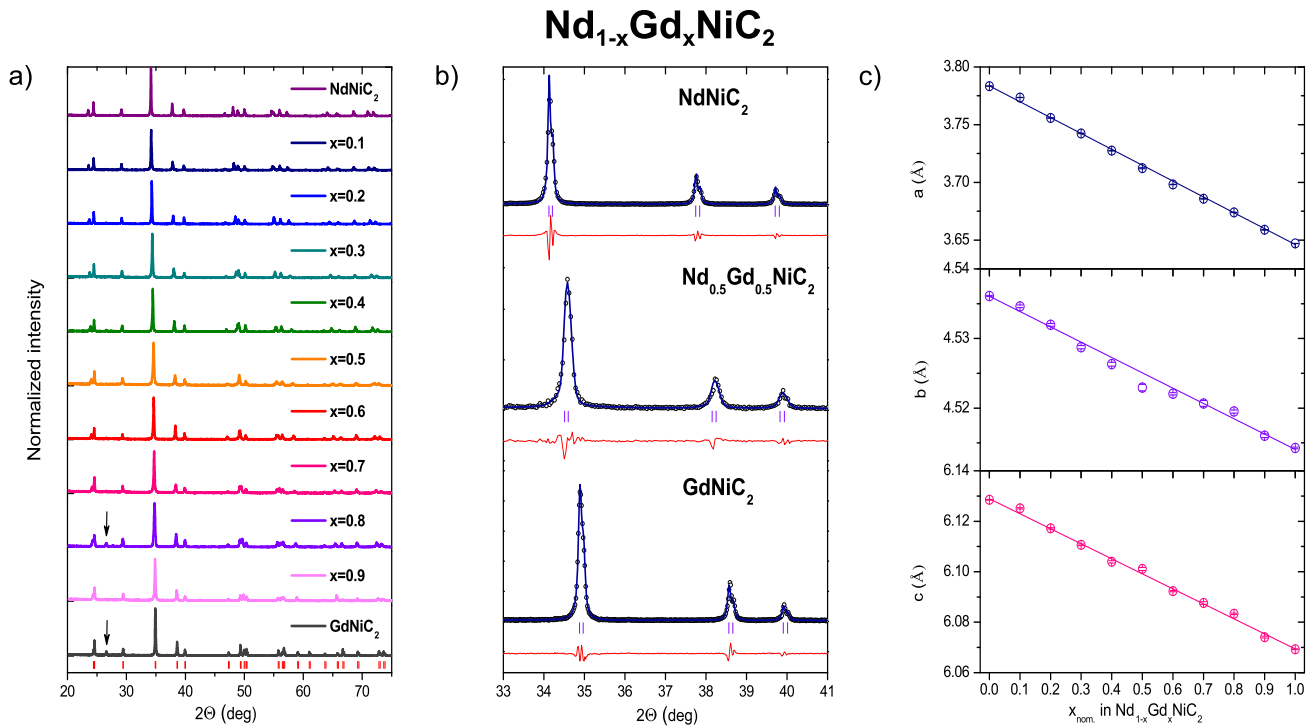}
 \caption{\label{XRD} a) Powder X-ray diffraction (pXRD) patterns for the series Nd$_{1-x}$Gd$_x$NiC$_2$ (0 \(\le\) x \(\le\) 1). The vertical ticks correspond to the Bragg peaks for Nd$_{1-x}$Gd$_x$NiC$_2$. Arrows indicate the peaks corresponding to residual carbon content. b) Expanded view of the main reflection line (111) showing a shift towards higher angles by substituting Gd for Nd. Open circles denote experimental points, whereas calculated diffraction patterns are represented by the solid blue lines. Differences between experiment and a model are shown by the red lines. c) Change of $a$, $b$ and $c$ lattice parameters for the series Nd$_{1-x}$Gd$_x$NiC$_2$ (0 \(\le\) x \(\le\) 1)
 }
  \end{figure*}
Overall loss of weight after the melting and annealing process was negligible (\(\le\)1\%) indicating that the elemental concentration was close to the actual alloying level. 
\\
\indent
For the crystal structure determination, powder X-ray diffraction (pXRD)measurements were performed using a X’Pert PRO-MPD, PANalitycal diffractometer with Cu K$_{\alpha}$ radiation, in the 2\(\theta\) range from 20$^o$ to 75$^o$. The lattice parameters were determined from a LeBail profile refinement of X-ray diffraction patterns for the entire Nd$_{1-x}$Gd$_x$NiC$_2$ series executed using FULLPROF software \cite{FULLPROF}.
\\
\indent
The physical property measurements were performed in the temperature range of 1.9-300 K by using a commercial Physical Property Measurement System (Quantum Design). Magnetization measurements were carried out using the AC and the DC Susceptibility Option (ACMS). A standard four-probe contact configuration was used to measure the electrical resistivity and the platinum wires (\(\phi\) = 37\(\mu\)m) were attached to the polished samples by spot welding.

\section{Results and discussion}
\indent
The phase composition and crystallographic structure of the obtained samples were checked at room temperature by powder X-ray diffraction which revealed that all observed reflections for the Nd$_{1-x}$Gd$_x$NiC$_2$ (0 \(\le\) x \(\le\) 1) series are indexed in the orthorhombic CeNiC$_2$-type structure with the space group $Amm2$. The pXRD patterns for Nd$_{1-x}$Gd$_x$NiC$_2$ solid solutions are presented in Fig. \ref{XRD} a). 
Only for the x=0.8 and x=1 samples, additional weak reflection lines (marked by arrows) corresponding to residual carbon content are observed. The substitution of Nd with Gd does not change the crystal structure symmetry. However, one can observe that the Bragg reflection lines are shifted towards higher angles with an increase in the Gd content (shown in Fig.\ref{XRD} b)). This behavior is consistent with Gd$^{3+}$ having a smaller ionic radius than Nd$^{3+}$ and confirms successful chemical alloying.
\\
\indent
 The unit cell parameters determined from LeBail refinement for the parent compounds NdNiC$_2$ and GdNiC$_2$   were found to be: $a$ = 3.783(1) Å, $b$ = 4.536(1) Å, $c$ = 6.129(1) Å and $a$ = 3.647(1) Å, $b$ = 4.514(1) Å, $c$ = 6.069(1) Å, respectively. These values are in good agreement with those reported in the literature \cite{Jeitschko1986}. The refined lattice parameters for the intermediate samples from the Nd$_{1-x}$Gd$_x$NiC$_2$ series  are shown in Fig.\ref{XRD} c). The $a$, $b$ and $c$ parameters decrease linearly with an increase in the Gd concentration for the whole x range, and hence obey Vegard's law. The $a$ constant expands by almost 4\% wheras the changes of the $b$ and $c$ parameters are less pronounced (below 1\%). The smallest change is observed for the $b$ parameter, which could be associated with rigid the C-C dimers along the $b$ axis\cite{Murase2004}. 
\begin{figure} [t!]
  \includegraphics[angle=0,width=1.05\columnwidth]{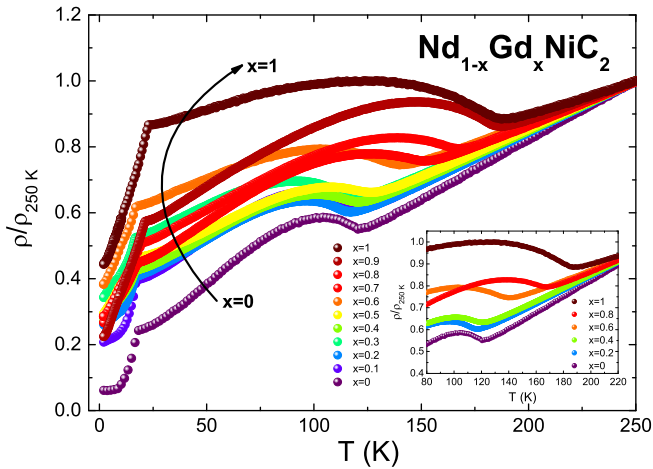}
 \caption{\label{opor} Temperature dependence of the normalized electrical resistivity $\rho/\rho_{250 K}$ (T) for Nd$_{1-x}$Gd$_x$NiC$_2$ (0 \(\le\) x \(\le\) 1). Inset shows the vicinity of the CDW transition for selected samples for better clarity.}
  \end{figure}

The electrical resistivity for the Nd$_{1-x}$Gd$_x$NiC$_2$ (0 \(\le\) x \(\le\) 1) series was measured without an applied magnetic field in the temperature range 1.9 – 250 K and the results ($\rho/\rho_{250 K}$ vs. T) are shown in Fig. \ref{opor}. The whole series exhibits typical metallic behavior at high temperatures showing a decrease of the electrical resistivity with decreasing temperature. With further cooling, a minimum followed by a hump well known to be a characteristic feature of a charge density wave transition, is observed for the entire concentration of Gd in the Nd$_{1-x}$Gd$_x$NiC$_2$ series. The temperature of the CDW formation ($T_{CDW}$) was obtained from the temperature derivative of the resistivity ($d\rho/dT$) and for the parent compounds NdNiC$_2$ and GdNiC$_2$, $T_{CDW}$ is 130 K and 197 K, respectively. The inset of Fig. \ref{opor}. shows the expanded view of the normalized electrical resistivity in the vicinity of the CDW transition for selected Nd$_{1-x}$Gd$_x$NiC$_2$ samples. With the increase in the  Gd concentration, the temperature of the CDW transition ($T_{CDW}$) for the Nd$_{1-x}$Gd$_x$NiC$_2$ series starts to decrease from 130 K for x = 0 (purple spheres), reaching a minimum of 123 K for the Gd concentration x = 0.2 (blue spheres) and then increases more rapidly with a further increase of Gd up to 197 K for x = 1 (brown spheres). Upon further cooling, the electrical resistivity for the whole series continues to decrease until the visible drop in resistivity at low temperatures. For GdNiC$_2$ and NdNiC$_2$ this effect has been reported to be caused by an  antiferromagnetic transition, and therefore it is reasonable to expect the same behavior for the intermediate compounds.

\begin{figure}[ht]
\includegraphics[angle=0,width=1.0\columnwidth]{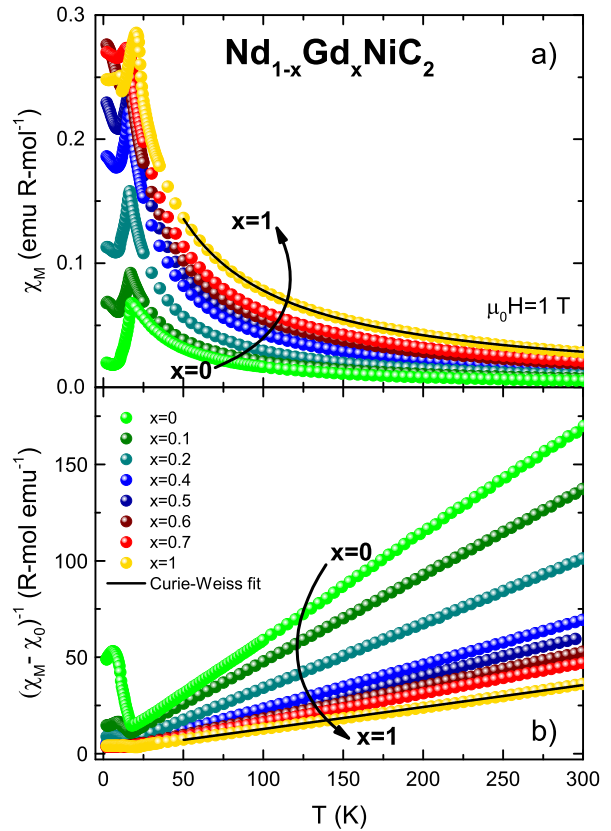}
 \caption{\label{chi} Temperature dependence of the molar magnetic susceptibility \(\chi\)$_M$ (a) and of the reciprocal molar magnetic susceptibility $(\chi\)$_M$-\(\chi\)$_0)^{-1}$ (b) for selected samples from the Nd$_{1-x}$Gd$_x$NiC$_2$ (0 \(\le\) x \(\le\) 1) series.}
  \end{figure}

\begin{figure}[t!]
\includegraphics[angle=0,width=1.0\columnwidth]{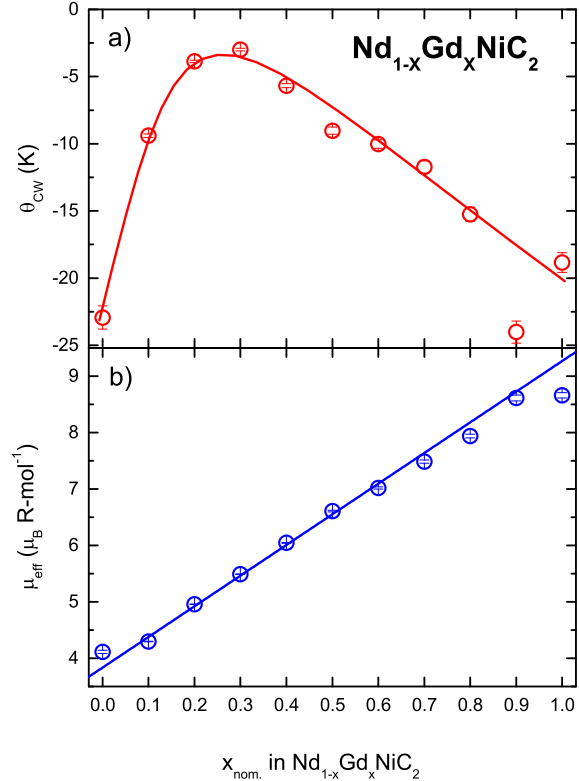}
 \caption{\label{TCWimieff} Change of the Curie-Weiss temperature \(\theta\)$_{CW}$ (a) and effective magnetic moment \(\mu\)$_{eff}$ (b) for Nd$_{1-x}$Gd$_x$NiC$_2$ (0 \(\le\) x \(\le\) 1). Solid lines are guide for the eye. }
  \end{figure}

The temperature dependence of the magnetic susceptibility \(\chi\)(T) for the Nd$_{1-x}$Gd$_x$NiC$_2$ (0 \(\le\) x \(\le\) 1) series measured with a \(\mu\)$_{0}$H = 1 T applied magnetic field is depicted in Fig. \ref{chi} a) (shown only for selected samples for better clarity). At high temperatures the entire series shows paramagnetic behavior. Between 16 K and 22 K  (depending on $x$), $\chi(T)$ reveals a sharp maximum. The Néel temperature ($T_N$) was estimated as the maximum of the temperature derivative of the magnetic susceptibility multiplied by the temperature (d(\(\chi\)T)/dT). The obtained $T_N$ values are in good agreement with those determined from the resistivity measurement. An additional minimum followed by a further increase is observed for most members from the Nd$_{1-x}$Gd$_x$NiC$_2$ series and can be attributed to a spin-flop transition as reported for the GdNiC$_2$ compound \cite{Hanasaki2011, Hanasaki2017}.
\\
\indent
Above $T_N$, the entire series obeys the Curie-Weiss law. The \(\chi\) (T) were fitted using the Curie-Weiss law expression: 
\\
\begin{equation}
\label{CurieWeiss}
\chi (T) = \frac{C}{T-\theta_{CW}} + \chi_0
\end{equation}
\\
where C is the Curie constant, \(\theta\)$_{CW}$ is the Curie-Weiss temperature and \(\chi\)$_0$ is the temperature independent magnetic susceptibility which is related both to the sample and the sample holder (a small diamagnetic contribution from sample straw). An exemplary fit to the data is shown with a solid line in Fig. \ref{chi}.
The results of the magnetic susceptibility with a clear magnetic anomaly at $T_N$ were also presented as a function of the reciprocal magnetic susceptibility with temperature $(\chi\)$_M$-\(\chi\)$_0)^{-1}$ vs. $T$ in Fig. \ref{chi} b). Above the AFM transition temperature, all $(\chi\)$_M$-\(\chi\)$_0)^{-1}$plots show an approximate linear dependence. 
\\
\indent
Having determined the value of the Curie constant C from the Curie-Weiss fit, the effective magnetic moment \(\mu\)$_{eff}$ was calculated for each compound of the Nd$_{1-x}$Gd$_x$NiC$_2$ series using the formula:
\\
\begin{equation}
\label{EffectiveMoment}
\mu_{eff} = \sqrt{\frac{3 C k_B}{{\mu_B}^2 N_A}}
\end{equation}
\\
where k$_B$ is the Boltzmann constant, N$_A$ is the Avogadro number and \(\mu\)$_B$ is the Bohr magneton. 
\\
\indent
The Curie-Weiss temperature and the effective magnetic moment versus  Gd concentration (\(\theta\)$_{CW} (x)$ and \(\mu\)$_{eff} (x))$ are presented in Fig. \ref{TCWimieff} a) and b), respectively.  Estimated \(\theta\)$_{CW}$  for GdNiC$_2$ denotes -18.85 K and stands in good agreement with previously reported values \cite{Matsuo1996}. The \(\theta\)$_{CW}$ =  -22.93 K obtained for NdNiC$_2$ is however visibly different from the value reported by us previously (-5.9 K) \cite{Kolincio2017}. The \(\theta\)$_{CW}$ in this compound has been found very sensitive to the direction of magnetic field and vary from -17.8 K along b axis to 24.6 along a axis \cite{Onodera1998}. The inconsistency with our last results can then be attributed to a difference in the samples microstructure. The negative sign of the Curie-Weiss temperature indicates  antiferromagnetic fluctuations. Upon the crossover from NdNiC$_2$ to GdNiC$_2$, the $\theta_{CW}$ initially shifts towards less negative values and reaches a maximum for the intermediate compound Nd$_{0.7}$Gd$_{0.3}$NiC$_2$ ($\theta_{CW}$= - 1.35 K). The proximity  to zero suggests the weakness of the magnetic interactions between magnetic ions. With a further increase of the Gd concentration, \(\theta\)$_{CW}$ becomes more negative again, which is a signature of the enhancement of antiferromagnetic interactions. For x = 0.9, a deviation from the curve is observed and the origin of this anomaly is unknown. 
\\
\indent
In the RNiC$_2$ family, nickel atoms do not contribute to the magnetic moment and the magnetic ordering originates only from the 4$f$ electrons of rare earth ions R$^{3+}$. 
The effective magnetic moments of the parent compounds determined from the Curie-Weiss fit, (\(\mu\)$_{eff}$ = 4.11\(\mu\)$_B$ and 8.66\(\mu\)$_B$  for NdNiC$_2$  and GdNiC$_2$, respectively) are larger than the values expected for free R$^{3+}$ ions (3.62\(\mu\)$_B$ for Nd$^{3+}$ and 7.94\(\mu\)$_B$ for Gd$^{3+}$) but close to the values reported previously\cite{Yakinthos1990, Matsuo1996}. The change of the effective magnetic moment with increasing level of the Gd concentration  \(\mu\)$_{eff}$(x) could be considered as linear with a small deviations for the parent compounds (NdNiC$_2$ and GdNiC$_2$). This result is consistent with what can be expected from electron introduction when Gd ($4f^7$) replaces Nd ($4f^3$) and the deviation from linearity could be caused by a disorder effect introduced by doping.
\\
\indent
The de Gennes law describing the strength of the indirect exchange coupling between local moments predicts that, for the systems in which the magnetic ground state originates from the RKKY interaction, the Néel temperature is expected to be related to the bulk electronic density of states at the Fermi level $N(\epsilon_F)$ with relationship:

\begin{equation}
\label{dGeq1}
T_N\sim8N(\epsilon_F)k_BI^2dG
\end{equation}
where $k_B$ is the Boltzmann constant and $I$ is the exchange integral. The de Gennes factor ($dG$) is given by the formula: 

\begin{equation}
\label{dGeq2}
dG=(g_J-1)^2J(J+1)
\end{equation}
where $g_J$ is the Landé factor and $J$ is the total angular momentum of the $R^{3+}$ ion following Hund's rule in the ground state.  The effective $dG$ factor for the Nd$_{1-x}$Gd$_x$NiC$_2$ solid solutions was calculated as a weighted average of the two elemental $dG$ factors:
\begin{equation}
\label{dGeq3}
dG_{eff} = (1-x) dG_{Nd} + (x) dG_{Gd}
\end{equation}

\begin{figure}[t!]
\begin{center}
\includegraphics[angle=0,width=1.05\columnwidth]{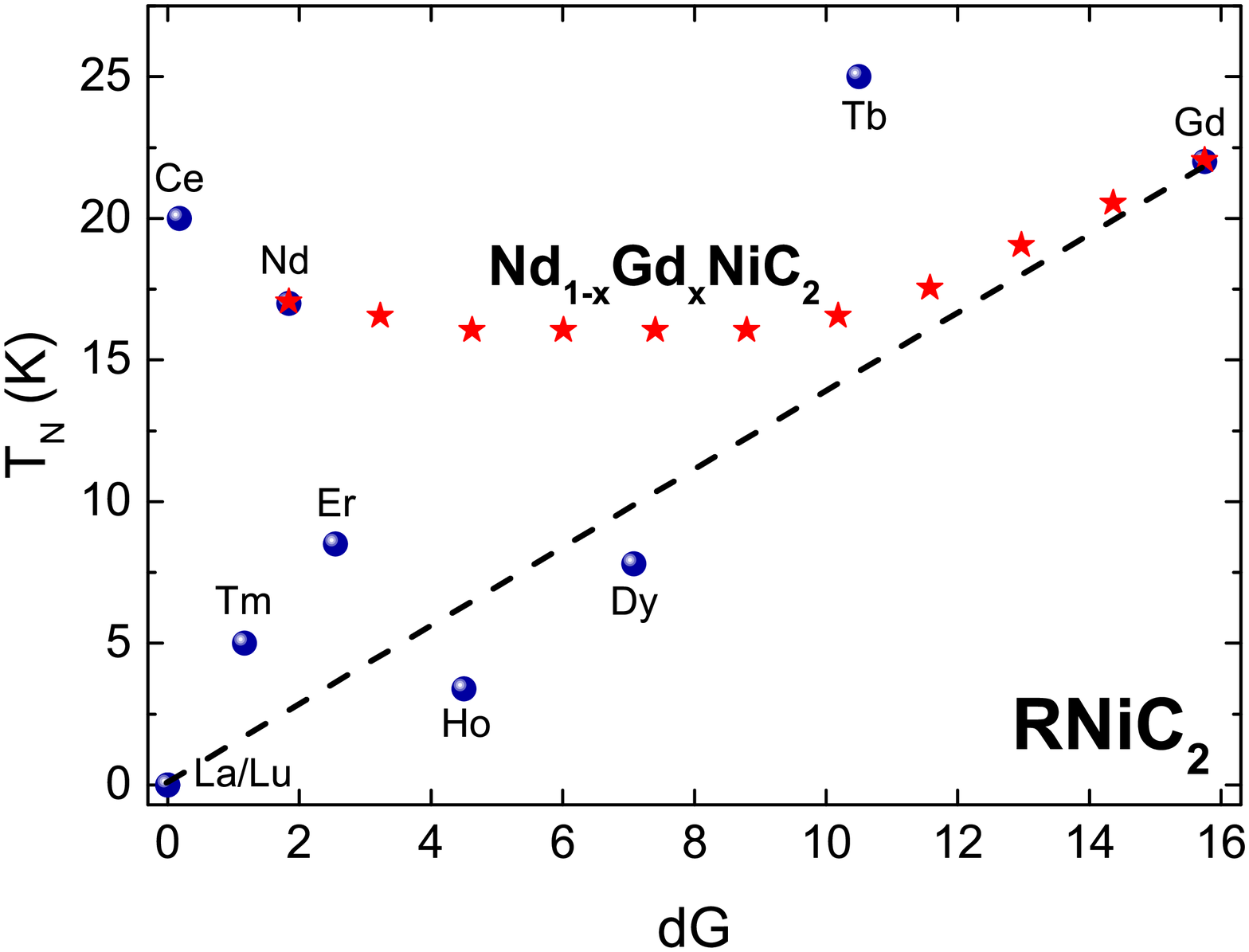}
\end{center}

 \caption{\label{DG}  The Néel temperature as a function of de Gennes factor for RNiC$_2$ family, including Nd$_{1-x}$Gd$_x$NiC$_2$ solid solution.}
  \end{figure}
  
Fig. \ref{DG} depicts de Gennes scaling for the members of the RNiC$_2$ family exhibiting an  AFM transition, including Nd$_{1-x}$Gd$_x$NiC$_2$ studied in this paper. A clear deviation from the expected de Gennes trend indicates that the bulk RKKY interaction is not essential to describe the magnetic transition in Nd$_{1-x}$Gd$_x$NiC$_2$ and other factors have to be considered. Previously, the breakdown of the $dG$ scaling for TbNiC$_2$ has been explained by the influence of the crystalline electric field (CEF)\cite{Koshikawa1997}. This scenario could be relevant in the case of Nd$_{1-x}$Gd$_x$NiC$_2$, since the deviation from the $dG$ scaling is visibly enhanced with an increase of the Nd content. This type of crossover can be expected based on the behavior of the parent compounds: GdNiC$_2$ shows negligible CEF\cite{Matsuo1996}, while the crystalline field plays a more important role in NdNiC$_2$ \cite{Onodera1998}. One must however find, that the violation of the de Gennes law being observed for NdNiC$_2$ is notably more pronounced than the deviations from the $dG$ scaling seen for Er and Tm bearing compounds. This observation stands in contrast with the comparison of the values of the CEF parameters $A_2^0$ and $A_2^2$ reported for these three compounds, which for NdNiC$_2$ are an order of magnitude lower than for ErNiC$_2$ and TmNiC$_2$ \cite{Koshikawa1997, Onodera1998}. For that reason, the alternative mechanisms have to be taken into account to explain this unusual effect.

According to equation \ref{dGeq2}, in the discussion of $T_N$ behavior one must also consider the role of the density of states, which is expected to be modified upon undergoing a Peierls transition inducing the opening of the electronic gap at the Fermi level and condensation of the free electronic carriers into the CDW state.

\begin{figure}[t!]
\includegraphics[angle=0,width=1.0\columnwidth]{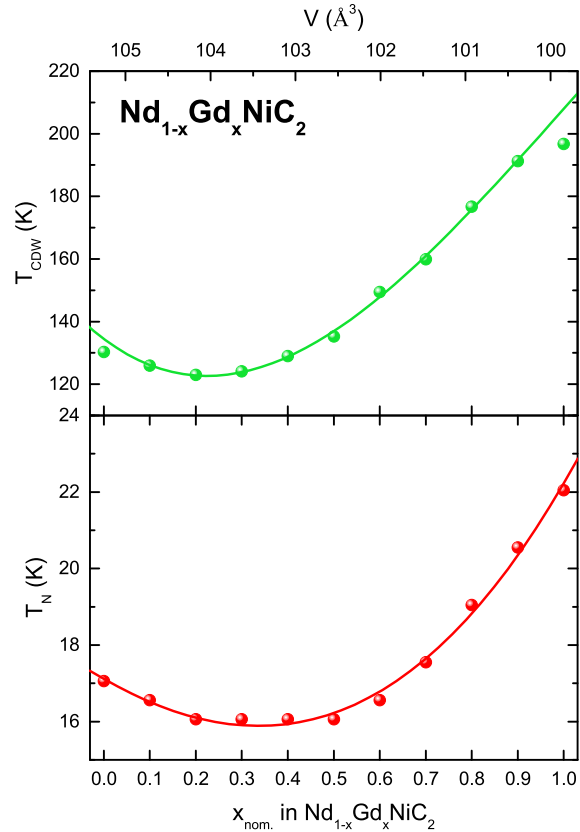}
 \caption{\label{TCDW} Evolution of $T_{CDW}$ (a) and $T_N$ (b) as a function of Gd concentration x$_{nom.}$ in the Nd$_{1-x}$Gd$_x$NiC$_2$ (0 \(\le\) x \(\le\) 1) }
  \end{figure}
In Fig. \ref{TCDW} we compare the CDW transition temperature $T_{CDW}$ (panel a) and the Néel temperature $T_N$ (panel b) plotted against the values of the Gd concentration in the Nd$_{1-x}$Gd$_x$NiC$_2$ (0 \(\le\) x \(\le\) 1).  For better clarity, the unit-cell volume is displayed above the top axis. The character of the evolution of these transition temperatures is similar and reminiscent of the behavior of $\theta_{CW}(x)$ - both curves reveal minimum with the composition corresponding to x = 0.3.

The correlation between the $T_{CDW}$ and $T_N$ suggests a strong interrelationship between the charge density wave state an antiferromagnetism. 
According to the de Gennes theory, one would expect a negative coupling, since the CDW transition decreases the $N(\epsilon_F)$, thus, according to eq. \ref{dGeq1}, the CDW should have a negative impact on the magnetic interactions. The stronger effect is expected to occur when the Peierls temperature is higher and the electronic gap is increased. In the mean field approach\cite{Monceau2012}, these quantities are correlated by:
\begin{equation}
\label{MF}
2\Delta=3.52k_BT_{CDW}
\end{equation}

Nevertheless, one should not underestimate the role played by the Fermi surface nesting vectors. As established from both theoretical predictions \cite{Roth1966, Bruno1992, Aristov1997, Aristov2000, Simon2008} and experimental results \cite{Inosov2009, Zhou2010, Feng2013}, in the case of a non-trivial topology of the Fermi surface, the RKKY interaction becomes sensitive to the delicate character of the nesting conditions. The direct link with the FS curvature makes the RKKY interaction strongly anisotropic and leads to the deviations from the simplistic isotropic approach expressed by equation \ref{dGeq1}. The common aspect of the FS nesting and momentum dependent RKKY interaction lies in the fact that both phenomena are associated with the generalized electron (spin) susceptibility represented by the  Lindhard function\cite{Roth1966, Bruno1992, Simon2008}:
\begin{equation}
\label{Lindhard}
\chi^0(q)\sim\sum\limits_{k}\frac{f_{k+q}-f_k}{\epsilon_{k+q}-\epsilon_k}
\end{equation}
where $f_k$ is the Fermi distribution function and  $\epsilon_k$ denotes for the energy corresponding to the state with wavevector $k$. The course, or more strictly, the maximum or a singularity of $\chi^0(q)$ leading to the nesting of the Fermi surface can significantly enhance the strength of the indirect interaction between the magnetic moments.
Simultaneously the Fermi surface nesting is a common feature associated with the formation of charge density waves \cite{Monceau2012, Johannes2008}. The same Lindhard function determines the energy gain from the electronic part of the CDW. 
Thus, this function often plays a decisive role for the preferred $q$ vector of the CDW modulation \cite{Rossnagel2011}, which in most CDW systems is identical with the FS nesting vector.
The anisotropic RKKY interaction is therefore significantly enhanced in the specific reciprocal space directions, when the magnetic propagation vector coincides with the values of $q$ corresponding to the maximum of $\chi^0(q)$, consistent with the CDW modulation. The experimental evidence for such nesting enhanced behavior has been reported for Gd$_2$PdSi$_3$, Tb$_2$PdSi$_3$ \cite{Inosov2009}, GdSi\cite{Feng2013}, Yttrium \cite{Dugdale1997} or Gd-Y alloys \cite{Fretwell1999}. 
 The CDW modulation vectors for NdNiC$_2$ and GdNiC$_2$ defined from X-ray diffuse scattering experiment, respectively $q_{Nd}$ = (0.5, 0.52, 0) \cite{Yamamoto2013} and $q_{Gd}$ = (0.5, 0.5, 0) \cite{Shimomura2016} have also been theoretically predicted as genuine FS nesting vectors\cite{Laverock2009}. These vectors stand in agreement with the wavevectors describing the AFM order (0.5, 0.5, 0) observed for NdNiC$_2$\cite{Yakinthos1990, Schafer1992} and proposed for GdNiC$_2$\cite{Matsuo1996}. It is reasonable to assume that this coincidence is relevant also for the solid solutions between NdNiC$_2$ and GdNiC$_2$, giving rise to an enhancement of the AFM order due to a cooperative effect with FS nesting accompanying the Peierls instability. The scenario of affirmative coupling between the CDW and magnetism in these systems is also supported by the recent work of Hanasaki et al. \cite{Hanasaki2017}, who suggested that the origin of the antiferromagnetic ground state in GdNiC$_2$ lies in the spin density wave constructed upon the preexisting CDW. In this model, the charge density modulated as a result of the Peierls instability is composed of two distinct spin-up and spin-down charge distributions and while the presence of strong magnetic moments produces a phase shift between them, the periodical spin density modulation is formed, giving rise to the enhancement of antiferromagnetic coupling between  local $f$ moments.

\begin{figure}[t!]
\includegraphics[angle=0,width=1\columnwidth]{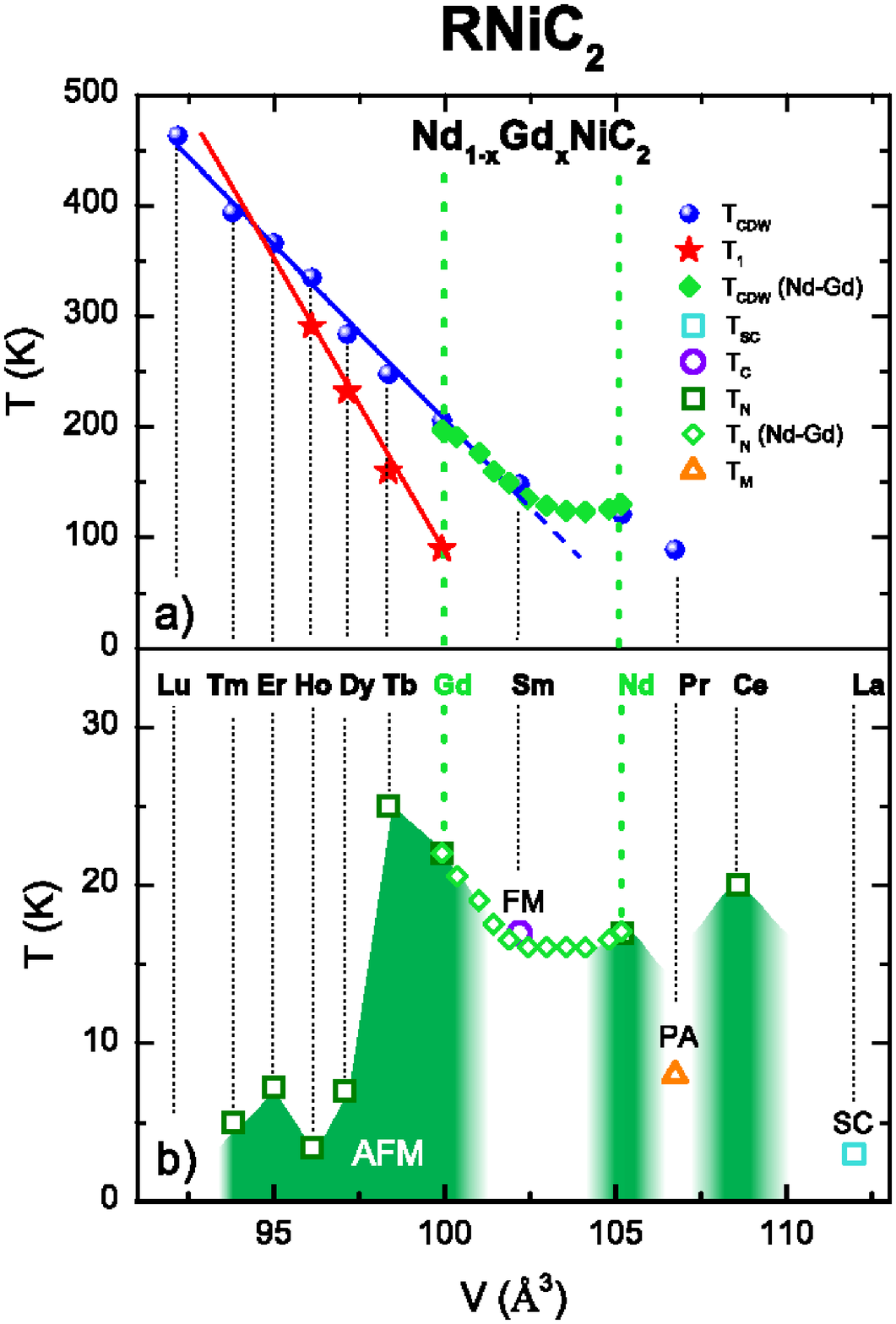}
 \caption{\label{diagram} Phase diagram of the RNiC$_2$ family including Nd$_{1-x}$Gd$_x$NiC$_2$ (0 \(\le\) x \(\le\) 1) solid solution. Upper panel (a) shows the variation of the Peierls ($T_{CDW}$) and lock-in ($T_1$) temperatures with the unit cell volume\cite{Lee1996, Kotsanidis1989, Onodera1998, Murase2004, Shimomura2016, Kolincio2017, Michor2017, Roman2018}. Lower panel (b) depicts the magnetic groundstates with their characteristic temperatures: $T_N$ (Néel), $T_C$ (Curie) and $T_M$ for the paramagnetic anomaly (PA) observed in PrNiC$_2$. $T_{SC}$ marks the onset of superconductivity for LaNiC$_2$.}
  \end{figure}

The values of $T_{CDW}$ and $T_N$ determined in this work have been imposed on the phase diagram of the RNiC$_2$ family, shown in Fig. \ref{diagram} as a bright green region. It can-not escape from the viewer's eye that, these results converge with the trend line $T_{CDW}(V)$ and $T_N(V)$ for RNiC$_2$. It is visible that near the point corresponding to SmNiC$_2$, the charge density wave temperature scaling starts to deviate from linearity. Our results (see Fig. \ref{chi}) reveal the AFM ground state of all the studied compounds from the Nd$_{1-x}$Gd$_x$NiC$_2$ series, even those in a close proximity to SmNiC$_2$, which is a ferromagnet. To confirm the genuine AFM character of the magnetic transitions of the Nd$_{1-x}$Gd$_x$NiC$_2$ series , we have measured the $M$ vs. $H$  (not shown here) indicating that no ferromagnetic ground state exists below $T_N$. Additionally, in contrast to SmNiC$_2$ which shows a rapid drop in resistivity below the magnetic transition temperature due to complete destruction of CDW and release of the electronic carriers \cite{Shimomura2009, Hanasaki2012, Lei2017}, a less abrupt decrease of the $\rho(T)$ curve is seen for Nd$_{1-x}$Gd$_x$NiC$_2$ series. The behavior of the solid solutions is reminiscent of the features reported for parent Nd and Gd bearing compounds. In NdNiC$_2$ and GdNiC$_2$, the CDW state partially survives the AFM transition and the similarity between the parent compounds and their solid solution suggests the identity of the observed mechanisms. 

\section{conclusions}
In this article we have examined the transport and magnetic properties of the Nd$_{1-x}$Gd$_x$NiC$_2$ (0 \(\le\) x \(\le\) 1) solid solution to explore the evolution of charge density wave and magnetism through the crossover from NdNiC$_2$ towards GdNiC$_2$. The variation of the Peierls temperature of Nd$_{1-x}$Gd$_x$NiC$_2$ as a function of the unit cell volume covers suitably the deviation from the linear trend observed in the previous study. We also report the breakdown of the de Gennes scaling in the studied series. The results are discussed in terms of the electric crystal field and indirect Ruderman-Kittel-Kasuya-Yosida interaction between local magnetic moments. The correlation between the Peierls, Néel and Curie-Weiss temperatures suggests a strong coupling between the Fermi surface nesting and the antiferromagnetic ground state, described by the compatible wavevectors. We also suggest that this hypothesis can be confirmed by angle-resolved photoemission spectroscopy (ARPES) experiment performed on single crystals.

\begin{acknowledgments}
 Authors gratefully acknowledge the financial support from National Science Centre (Poland), grant number:  UMO-2015/19/B/ST3/03127.
 \end{acknowledgments}

%

\end{document}